\documentclass[longauth]{aa}
\usepackage{graphicx}
\usepackage{natbib}

\usepackage{txfonts}
\usepackage{color}

\usepackage[colorlinks=true, pdfstartview=FitV, linkcolor=blue, citecolor=blue, urlcolor=blue, breaklinks=true]{hyperref} 

\newcommand{\Msolar}{\mbox{$\vec{M}_{\odot}\,$}}

\newcommand{\kms}{\mbox{$\mbox{km\,s}^{-1}$}\,}

\begin{document}

\title{The orbital parameters of the $\delta$ Cep inner binary system determined using 2019 HARPS-N spectroscopic data} 
\titlerunning{{The orbital parameters of the $\delta$ Cep inner binary system}}
\authorrunning{Nardetto et al. }
\author{N.~Nardetto \inst{1}
\and V.~Hocd\'e \inst{2} 
\and P.~Kervella \inst{3}
\and A.~Gallenne\inst{4} 
\and W.~Gieren\inst{5,6}  
\and D.~Graczyk \inst{2} 
\and A.~Merand \inst{7}
\and M.~Rainer \inst{8} 
\and J.~Storm \inst{9}  
\and G.~Pietrzy\'nski \inst{2}
\and B.~Pilecki \inst{2} 
\and E.~Poretti \inst{8}  
\and M.~Bailleul \inst{1} 
\and G.~Bras \inst{3}
\and A.~Afanasiev \inst{3}
}
\institute{Universit\'e C\^ote d'Azur, Observatoire de la C\^ote d'Azur, CNRS, Laboratoire Lagrange, France,Nicolas.Nardetto@oca.eu
\and Nicolaus Copernicus Astronomical Center, Polish Academy of Sciences, ul. Bartycka 18, PL-00-716 Warszawa, Poland
\and LESIA (UMR 8109), Observatoire de Paris, PSL, CNRS, UPMC, Univ. Paris-Diderot, 5 place Jules Janssen, 92195 Meudon, France 
\and Instituto de Astrofísica, Departamento de Ciencias Físicas, Facultad de Ciencias Exactas, Universidad Andrés Bello, Fernández Concha 700, Las Condes, Santiago, Chile
\and Universidad de Concepci\'on, Departamento de Astronom\'ia, Casilla 160-C, Concepci\'on, Chile
\and Unidad Mixta Internacional Franco-Chilena de Astronom\'ia (CNRS UMI 3386), Departamento de Astronom\'ia, Universidad de Chile, Camino El Observatorio 1515, Las Condes, Santiago, Chile
\and European Southern Observatory, Alonso de C\'ordova 3107, Casilla 19001, Santiago 19, Chile 
\and  INAF -- Osservatorio Astronomico di Brera, Via E. Bianchi 46, 23807 Merate (LC), Italy  
\and Leibniz Institute for Astrophysics, An der Sternwarte 16, 14482, Potsdam, Germany 
}

\date{Received ... ; accepted ...}

\abstract{An inner companion has recently been discovered orbiting the prototype of classical Cepheids, $\delta$ Cep,  whose orbital parameters are still not fully constrained.}
{We collected new precise radial velocity measurements of $\delta$~Cep in 2019 using the HARPS-N spectrograph
mounted at the  Telescopio Nazionale {\it Galileo}. Using these radial velocity measurements, we aimed to improve the orbital parameters of the system.} 
{We considered a template available in the literature as a reference for the radial velocity curve of the pulsation of the star. We then calculated the residuals between our global dataset (composed of the new 2019 observations plus data from the literature) and the template as a function of the pulsation phase and the barycentric Julian date. This provides the orbital velocity of the Cepheid component. Using a Bayesian tool, we derived the orbital parameters of the system.} 
{Considering priors based on already published \textit{Gaia} constraints, we find for the orbital period a maximum a posteriori probability of $P_\mathrm{orb}=9.32_{-0.04}^{+0.03}$ years (uncertainties correspond to the 95\% highest density probability interval), and we obtain an eccentricity $e=0.71_{-0.02}^{+0.02}$, a semimajor axis $a=0.029_{-0.003}^{+0.002}$ arcsecond, and a  center-of-mass velocity $V_{0}=-17.28_{-0.08}^{+0.08}$ \kms, among other parameters.} 
{In this short analysis we derive the orbital parameters of the  $\delta$ Cep inner binary system and provide a cleaned radial velocity curve of the pulsation of the star, which will be used to study its Baade-Wesselink projection factor in a future publication.}
\maketitle

\section{Introduction}\label{s_Introduction}

Delta Cep is the prototype of classical Cepheid variable stars. Its variability was discovered by \citet{good1786}, and to this day the star remains a cornerstone for the calibration of the distance scale. It is in particular a benchmark star for the calibration of the projection factor \citep{nardetto04, nardetto06b, nardetto17}, a physical quantity that plays a central role in the Baade-Wesselink method \citep{baade26, wesselink46} of distance determination \citep{storm11a, storm11b, trahin21}. $\delta$ Cep was recently discovered to be a spectroscopic binary \citep{anderson15}. Before, the star was thought to be a visual binary \citep {fernie66}.

Because of the configuration of the system in 2015 when their spectroscopic data were secured with the High Accuracy Radial velocity Planet Searcher for the Northern hemisphere (HARPS-N) instrument, \citet{nardetto17} did not detect any evidence of an inner companion. However, in 2019, simultaneous data with the HARPS-N and GIANO instruments were secured for a set of five Cepheids, including $\delta$ Cep, using the GIARPS (GIAno \& haRPS) mode \citep{claudi16}. Two studies focused on the He~I~10830 \AA\ spectral line and the effective temperature determination of Cepheids, respectively \citep{andri23, kov23}. In this Letter we report the 2019 HARPS-N radial velocity measurements of $\delta$~Cep and present clear evidence of the presence of an inner companion (Sect. \ref{s_HARPS-N}). We used these data, together with data from the literature  (Sect. \ref{s_data}), to derive the orbital parameters of the system using a Bayesian approach (Sect. \ref{s_method}). The results presented in Sect. \ref{s_results} include a cleaned radial velocity curve of $\delta$~Cep, with the companion removed, that will be used in a forthcoming publication. 
                                                   
\section{HARPS-N spectroscopic observations of $\delta$ Cep}\label{s_HARPS-N}

We secured 24 HARPS-N spectroscopic measurements of $\delta$~Cep from 14 June to 21 September 2019. HARPS-N is the northern hemisphere counterpart of the HARPS instrument installed at the ESO 3.6 m telescope at La Silla Observatory in Chile \citep{co12}. The instrument covers the wavelength range from 3800 to 6900 angstrom with a resolving power of $R \simeq 115 000$. The data span 13 cycles of pulsation from the first to last epoch. $\delta$ Cep shows secular period changes, as shown in the O-C diagram of \citet{Cso22}. We thus used the ephemeris from \citet{Cso22} to calculate the pulsation phase for each individual observation: $T_0=2412028.256$~d, $P_\mathrm{Puls}=5.3663671$~d, and $\frac{dP}{dt}=-1.05379*10^{-6}$ days yr$^{-1}$. The final products of the HARPS-N data reduction software installed at the Telescopio Nazionale {\it Galileo} (online mode) are background-subtracted, cosmic-corrected, flat-fielded, and wavelength-calibrated spectra (with and without merging of the spectral orders). To calculate the cross-correlated velocity, we used the \emph{iSpec} tool with a G2V template \citep{bc14, bc19}. As discussed in \citet{nardetto23}, there is no difference in the derived radial velocities when using a G2V or an F6I template. We then applied a Gaussian fit to the cross-correlated function to derive the radial velocity ($RV_\mathrm{cc-g}$) and its uncertainty. The results are presented in Table~\ref{Tab_log} and in the left panel of Fig. \ref{fig_wocor} (see the blue triangles). We clearly see a dispersion in the radial velocity measurements from 1994 to 2019, and we show in this Letter that this is due to the presence of a companion.

\begin{table}
\begin{center}
\caption{HARPS-N RV$\mathrm{cc-g}$ radial velocities of $\delta$~Cep.} \label{Tab_log}
\setlength{\doublerulesep}{\arrayrulewidth}
 \tiny
\begin{tabular}{lcrrccc|}
\hline \hline \noalign{\smallskip}
 BJD        &       $\phi$  &   cycle &     RV$_\mathrm{cc-g}$      &       $\sigma_\mathrm{RV_\mathrm{cc-g}}$      &       v$_\mathrm{orb}$    \\
\hline
 2458678.6286 & 0.21 & 0 & -16.88 & 0.09 & -20.94 \\
 2458679.6868 & 0.41 & 0 & -6.67 & 0.05 & -20.91 \\
 2458680.6928 & 0.59 & 0 & 3.64 & 0.05 & -20.88 \\
 2458681.5945 & 0.76 & 0 & 12.03 & 0.06 & -20.85 \\
 2458682.6092 & 0.95 & 0 & 20.27 & 0.10 & -20.82 \\
 2458683.6867 & 0.15 & 1 & -18.95 & 0.11 & -20.79 \\
 2458703.6634 & 0.88 & 4 & 18.83 & 0.08 & -20.05 \\
 2458713.7034 & 0.75 & 6 & 11.06 & 0.06 & -19.64 \\
 2458716.5982 & 0.29 & 7 & -13.26 & 0.08 & -19.52 \\
 2458718.5278 & 0.65 & 7 & 6.19 & 0.05 & -19.44 \\
 2458720.5207 & 0.02 & 7 & 3.21 & 0.09 & -19.36 \\
 2458721.7128 & 0.24 & 8 & -15.47 & 0.08 & -19.31 \\
 2458722.6852 & 0.42 & 8 & -6.26 & 0.05 & -19.27 \\
 2458727.5688 & 0.33 & 9 & -11.00 & 0.06 & -19.07 \\
 2458728.4818 & 0.50 & 9 & -1.78 & 0.04 & -19.03 \\
 2458734.5037 & 0.62 & 10 & 5.07 & 0.05 & -18.79 \\
 2458735.6787 & 0.84 & 10 & 16.90 & 0.07 & -18.75 \\
 2458736.6173 & 0.02 & 10 & 3.65 & 0.10 & -18.71 \\
 2458737.4097 & 0.16 & 11 & -18.69 & 0.10 & -18.68 \\
 2458738.6240 & 0.39 & 11 & -7.76 & 0.05 & -18.63 \\
 2458739.5063 & 0.55 & 11 & 1.39 & 0.04 & -18.60 \\
 2458740.4791 & 0.74 & 11 & 10.79 & 0.06 & -18.56 \\
 2458741.4684 & 0.92 & 11 & 21.01 & 0.09 & -18.53 \\
 2458747.5311 & 0.05 & 12 & -8.51 & 0.10 & -18.31 \\

\hline
  days  &  phase   &  Nbr   & \kms   &  \kms     &   \kms   \\
\hline 
\end{tabular}
\end{center}
Notes: BJD is the barycentric Julian date, $\phi$ is the pulsation phase of the observations determined using the ephemeris provided by \citet{Cso22}, "cycle" corresponds to the number of cycles since the first observation in this sample,  RV$_\mathrm{cc-g}$ is the cross-correlated radial velocity using the Gaussian fit of the cross-correlated function, $\sigma_\mathrm{RV_\mathrm{cc-g}}$ is the corresponding uncertainty, and v$_\mathrm{orb}$ is the orbital velocity correction that has been applied to RV$_\mathrm{cc-g}$ (see Sect. \ref{s_results}).
\end{table}

\begin{table}
\begin{center}
\caption{ MAP and 95\% HDPIs of the orbital parameters of the $\delta$ Cep SB1 binary system as derived from the Bayesian analysis. The reference in decimal year, $T$, corresponds to a BJD of $2445104.090$ days.} \label{Tab_res}
\begin{tabular}{ll}
\hline \hline 
parameter & value    \\
\hline
$P_\mathrm{orb}$ $[yr]$  &    $9.32_{-0.04}^{+0.03}$\\
$T$ $[yr]$   & $1982.294_{-0.101}^{+0.111}$\\
$e$     &  $0.71_{-0.02}^{+0.02}$\\
$a$ $["]$   &    $0.029_{-0.003}^{+0.002}$\\
$\omega$ [°]  &  $230_{-3}^{+4}$\\
$\Omega$ [°]  &  $78_{-50}^{+56}$\\
$i$ [°]  &  $124_{-12}^{+17}$\\
$V_{0}$ $[km/s]$ &   $-17.28_{-0.08}^{+0.08}$\\
 $\pi$ $[mas]$  &    $3.66_{-0.10}^{+0.09}$\\
$f/\pi$ $[pc]$   &  $27_{-3}^{+7}$\\
 $m_1$ $[M_\odot]$  &    $5.26_{-1.40}^{+1.26}$\\
$q$   &   $0.11_{-0.02}^{+0.03}$\\
\hline 
\end{tabular}
\end{center}
\end{table}

\section{Spectroscopic data in the literature}\label{s_data}

\begin{figure*}[htbp]
\begin{center}
\resizebox{0.33\hsize}{!}{\includegraphics[clip=true]{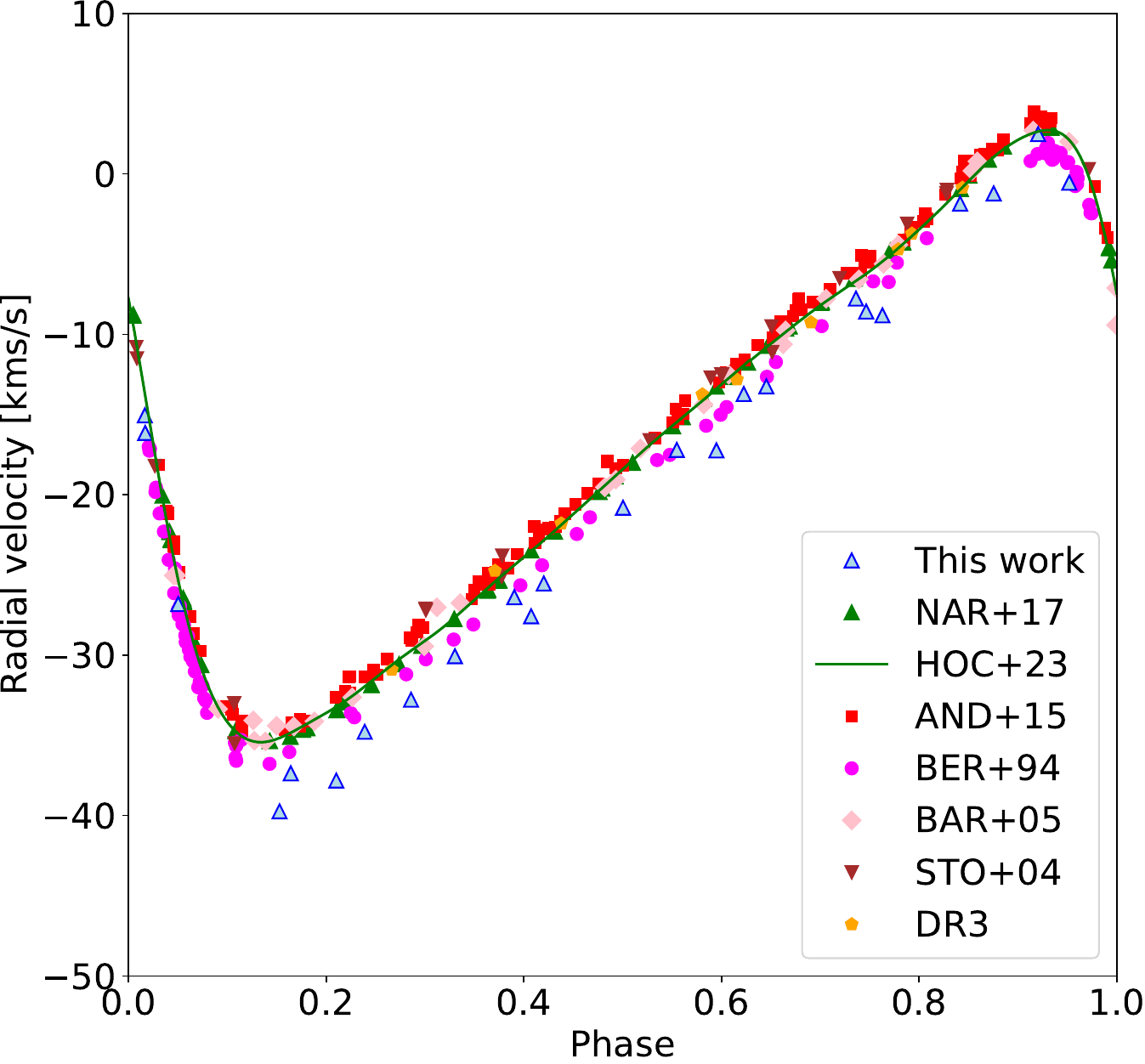}}
\resizebox{0.33\hsize}{!}{\includegraphics[clip=true]{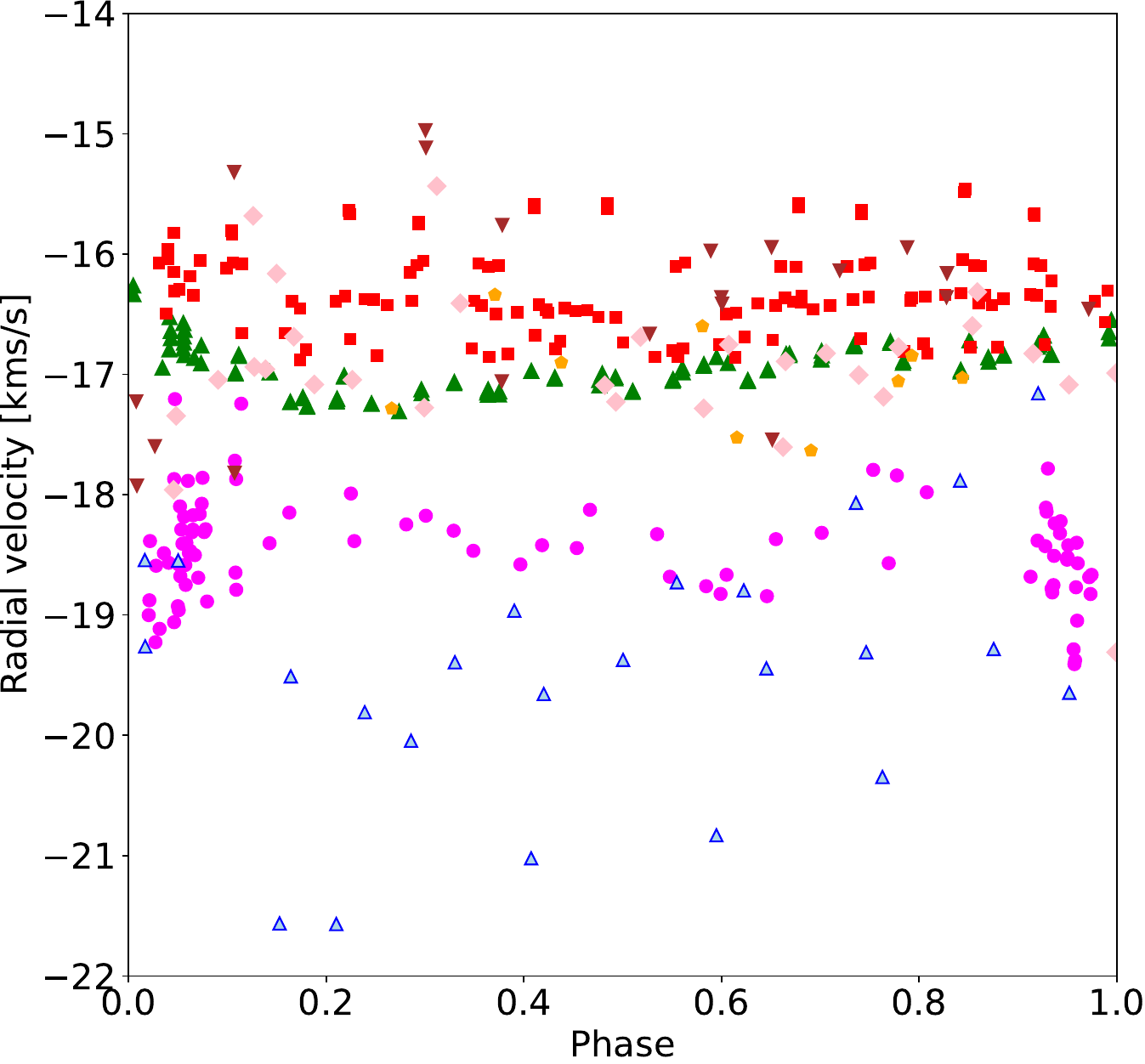}}
\resizebox{0.32\hsize}{!}{\includegraphics[clip=true]{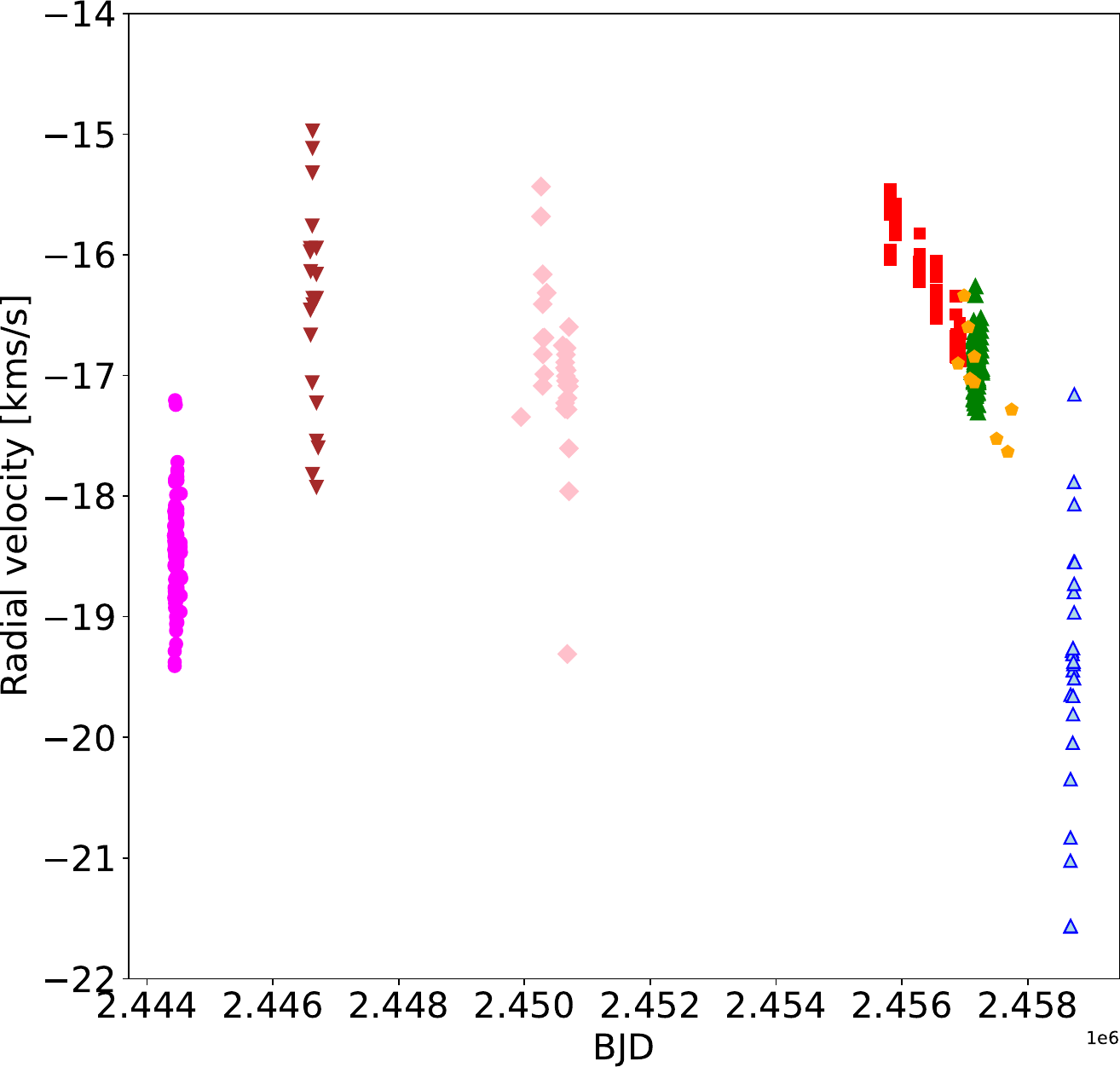}}
\end{center}
\caption{Radial velocity curves of $\delta$ Cep without correction from the presence of the companion. {\bf Left}: Comparison of the high-quality cross-correlated radial velocity curves, RV$\mathrm{cc-g}$ (Gaussian fit of the cross-correlated function), of $\delta$ Cep available in the literature. These data are not corrected for the center-of-mass velocity variation due to the inner companion, which explains the dispersion obtained in the curves. The data studied in this work are shown with blue triangles.  For comparison, the pulsation template provided by \citet{hocde23a} (green curve) has been shifted by the $\gamma$ velocity of -16.95 \kms found by \citet{nardetto17}. {\bf Middle}: Radial velocity curve from the left panel compared to the pulsation template. {\bf Right}: Same as the middle panel but as a function of the BJD. The long-term velocity variation due to the companion is clearly seen.}
\label{fig_wocor}
\end{figure*}

To derive the orbital parameters of the SB1 binary system composed of $\delta$ Cep and its inner companion, we needed to first extract the long-term orbital radial velocity of the system, that is to say, we had to remove the pulsation motion from the individual radial velocity measurements. For this, we considered the best-quality data in the literature, from \citet{bersier94b}, \citet{storm04b}, \citet{barnes05}, \citet{anderson15}, and \citet{nardetto17}, as well as data from \textit{Gaia} Data Release 3 \citep[DR3;][]{gaia16, gaia23}. The radial velocity measurements are presented in the left panel of Fig.~\ref{fig_wocor}. The next step was to subtract the  pulsation velocity from all these data in order to extract the orbital radial velocity of the Cepheid component.  In this work, we used the radial velocity curve template provided by \citet{hocde23a} based on the data from \citet{anderson15} as a reference for the radial velocity associated with the pulsation motion of the star (see the solid green line in the left panel of Fig.~\ref{fig_wocor}). This template by definition has a $\gamma$ velocity (i.e., an average value) of zero. \citet{nardetto17} investigated the possible effect of the binary motion due to the companion as discovered earlier by \citet{anderson15} but concluded that including a linear trend of  $-0.5 \pm 0.1$ m s$^{-1}$ d$^{-1}$ did not significantly reduce the residual in their measurements, which is of 0.5 m s$^{-1}$).  
Disentangling the $\gamma$ velocity of the Cepheid
and the receding or approaching motion of the center-of-mass velocity of the system for a given dataset is not simple, in particular when considering that the granulation of the star can affect the $\gamma$ velocity, as shown for the first time by \citet{nardetto08a} and \citet{vasilyev17, vasilyev18}. We see in Fig.~\ref{fig_wocor} that the template from \citet{hocde23a}, shifted by the $\gamma$ velocity value of $-16.95 \pm 0.005$ \kms found by \citet{nardetto17}, is indeed very close to the radial velocity curve obtained by \citet[see the green triangles]{nardetto17}. 


The residual between the data in the literature and our template is plotted as a function of the pulsation phase and barycentric Julian date (BJD) in the middle and right panels of  Fig.~\ref{fig_wocor}, respectively.  As discussed in \citet{anderson15}, there are some systematical velocity offsets between the instruments used by the different authors ($\sim$0.3 \kms at most), and possibly also drifts with time ($\sim$ 0.02 \kms), but they are difficult to determine for each instrument and are not free of errors. In this study we decided not to take them into account. We instead conducted some tests and show that such offsets have a negligible impact on our orbital parameter solution (see Sect~\ref{s_results}). 

\begin{figure*}[htbp]
\begin{center}
\resizebox{0.33\hsize}{!}{\includegraphics[clip=true]{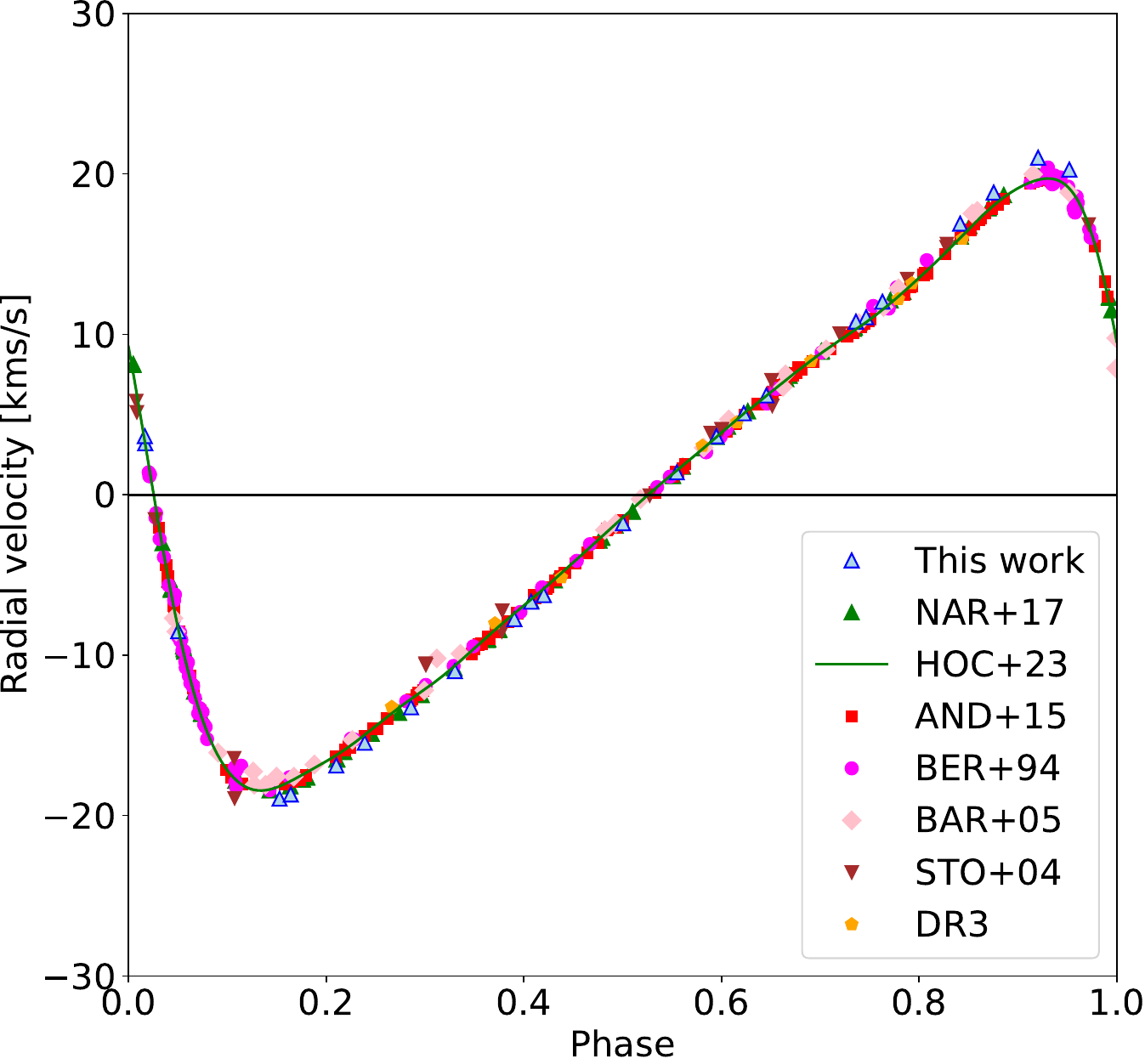}}
\resizebox{0.33\hsize}{!}{\includegraphics[clip=true]{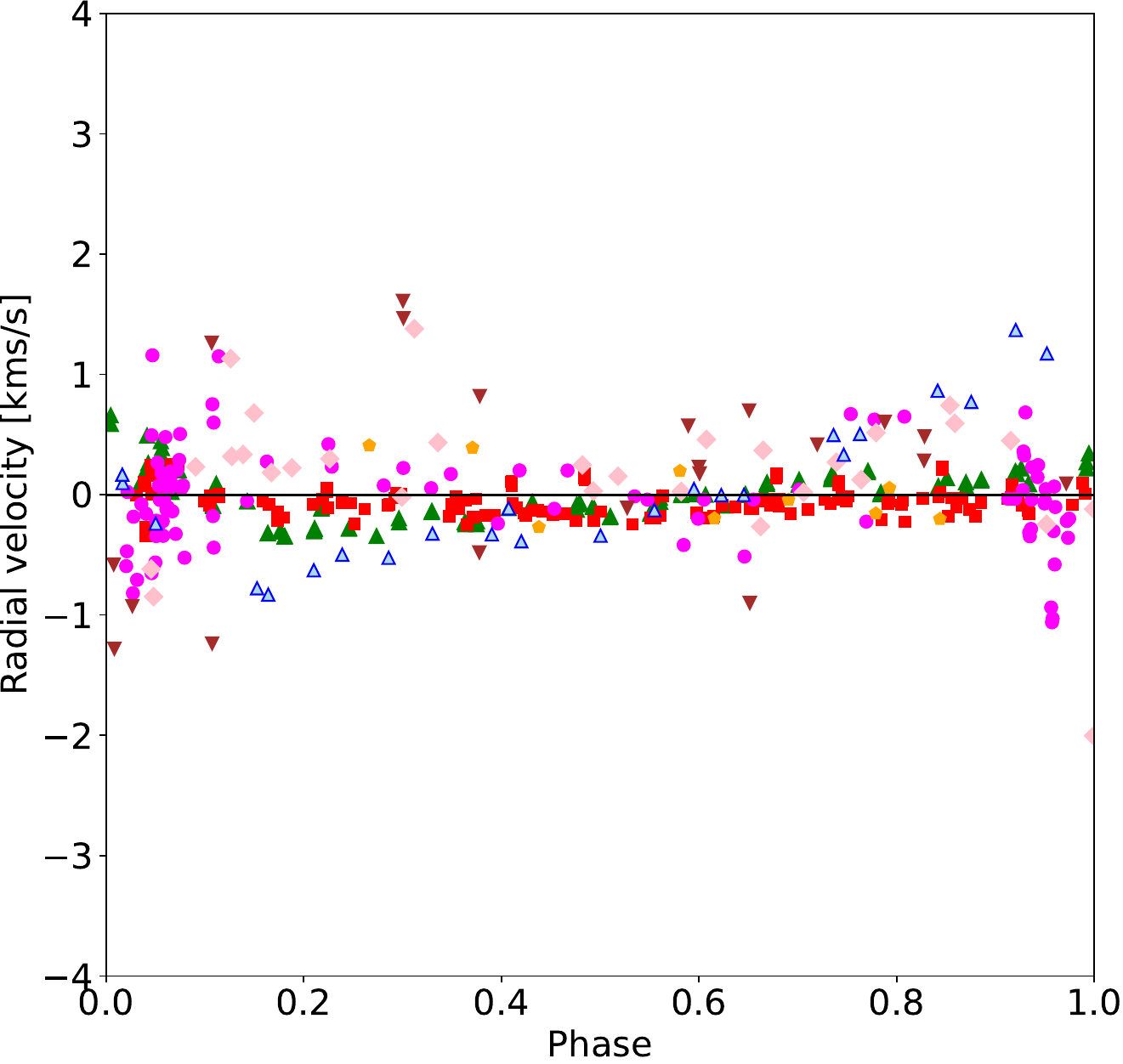}}
\resizebox{0.32\hsize}{!}{\includegraphics[clip=true]{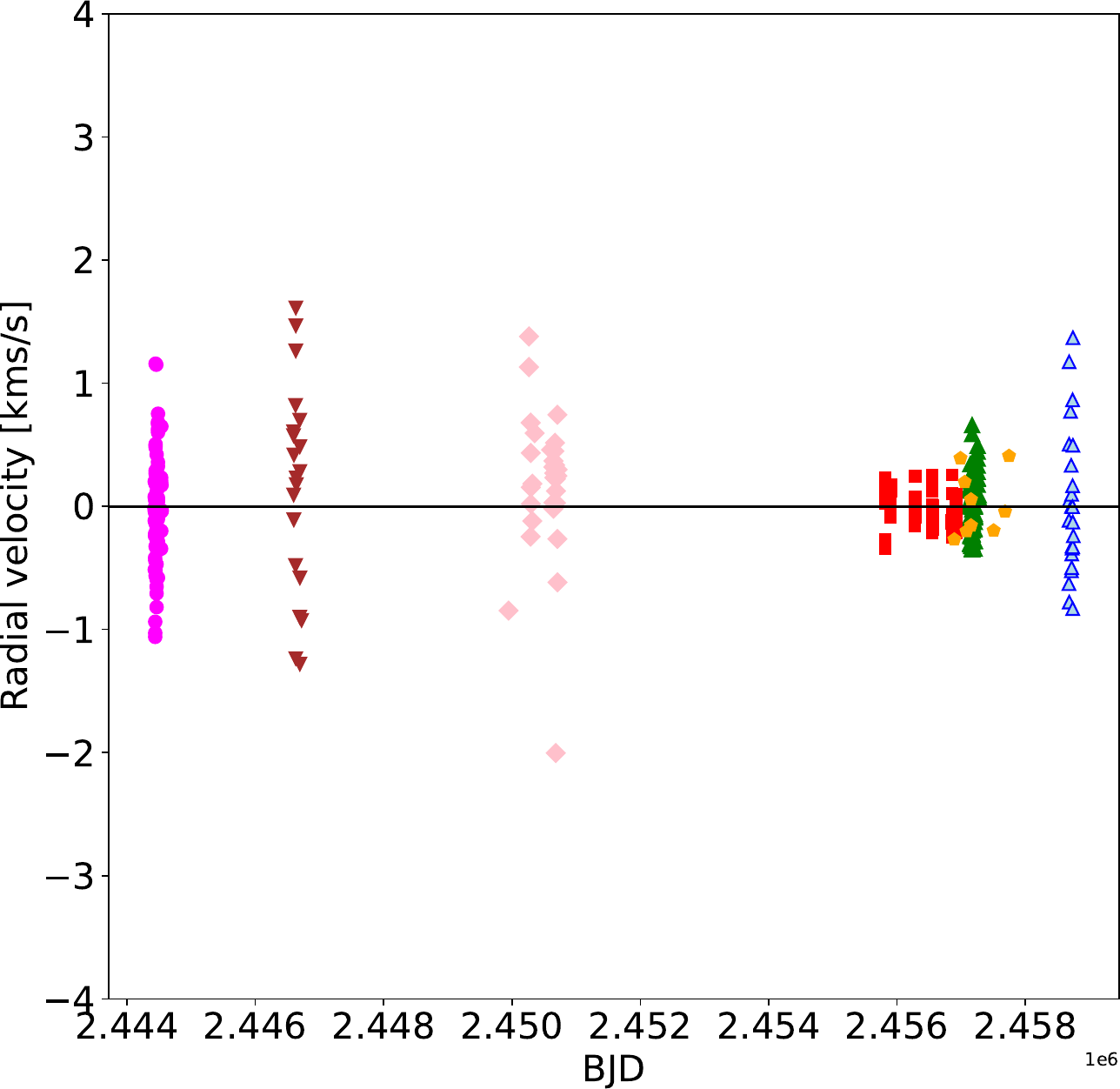}}
\end{center}
\caption{Same as Fig. \ref{fig_wocor} but after correction for the orbital velocity of the Cepheid component as described by the parameters listed in Table \ref{Tab_res}. The residuals in the middle and right panels are plotted as a function of the pulsation phase and BJD, respectively. The derived pulsation velocity curves shown in the left panel, including that from this work (blue open diamonds), have a dispersion much lower than in the left panel of Fig.~\ref{fig_wocor}. The rms residuals in the middle and right panels are about 0.4 \kms. The RV$\mathrm{cc-g}$ curve corrected for the binarity and corresponding to this work is also shown in Fig. \ref{fig_final}. }
\label{fig_wicor}
\end{figure*}

\section{A Bayesian approach to deriving the orbital parameters of the system}\label{s_method}

We applied a Bayesian inference methodology for the estimation of the orbital parameters to the single-line spectroscopic observations of $\delta$ Cep, based on the No-U-Turn sampler Markov chain Monte Carlo algorithm. For this, we used the BinaryStar tool available on GitHub\footnote{\url{https://github.com/mvidela31/BinaryStars}} \citep{carpenter17, videla22}. This tool is designed to provide a precise and efficient estimation of the joint posterior distribution of the orbital parameters in the presence of partial and heterogeneous observations. The tool allows to directly incorporate prior informations on the system. We defined five priors. First, we used the trigonometric parallax of the companion of $\delta$~Cep as found by \citet{kervella19a}:  $\pi = 3.364 \pm 0.049$ mas. The parallax from \textit{Gaia} DR3,  $\pi = 3.5551 \pm 0.1475$ mas \citep{gaia23}, has a renormalized unit weight error  of 2.7, indicating that it is not reliable. We also used four non-spectroscopic parameters from \citet[see their Table 2]{kervella19a} based on \textit{Gaia}: the estimation of the mass of the primary and secondary components,  $m_1=4.8 \pm 0.72$~\Msolar and $m_2=0.72 \pm 0.11$~\Msolar, the orbital inclination, $i=163\pm14$ deg, and the longitude of the ascending node, $\Omega=83\pm27$ deg. To ensure a Bayesian fit, we assumed a homogeneous uncertainty for all the radial velocity values in all datasets of 0.15 \kms, which corresponds to the average of all the available uncertainties.  

The outputs of our model are $\pi$, $i$, $\Omega,$ the time of periastron passage ($T$), the orbital period ($P_\mathrm{orb}$), the orbital eccentricity ($e$), the orbital semimajor axis ($a$), the argument of periapsis ($\omega$), the mass ratio of the individual components ($q =\frac{m_2}{m_1}$),  the velocity of the center of mass ($V_0$), and $\frac{f}{\pi}$, where $f=\frac{q}{1+q}$ is the fractional mass of the system. We ran different tests. We started by considering only two priors ($\pi$ and $m_1$), but  $\Omega$, $i$, $\frac{f}{\pi}$, and $q$ were poorly constrained in this case. Thus, we decided to take all five priors into account \citep{kervella19a}.  Using two or five priors does not significantly change the results regarding $T$, $P_\mathrm{orb}$, $e$, $a$, or $\omega$. 

\section{Results and discussion}\label{s_results}

\begin{figure*}[htbp]
\begin{center}
\resizebox{1.0\hsize}{!}{\includegraphics[clip=true]{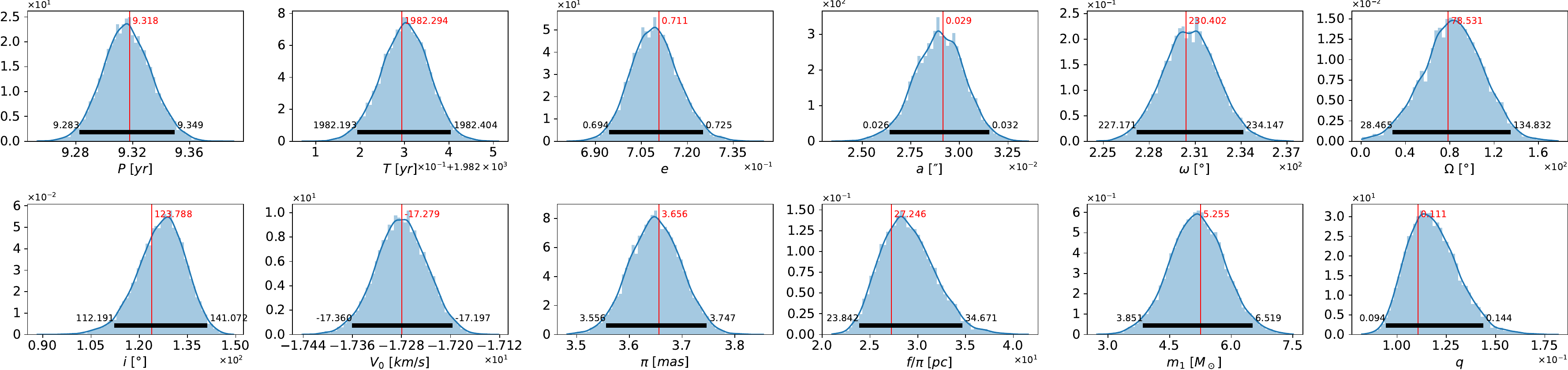}}
\end{center}
\caption{Marginal posterior distribution of the orbital parameters of the $\delta$ Cep system (SB1). The vertical line and the values in red correspond to the MAP of the Bayesian inference in the multi-parameter space, and the horizontal black line corresponds to the 95\% HDPIs. These values are listed in Table \ref{Tab_res}.}
\label{fig_MAP}
\end{figure*}

\begin{figure*}[htbp]
\begin{center}
\resizebox{0.95\hsize}{!}{\includegraphics[clip=true]{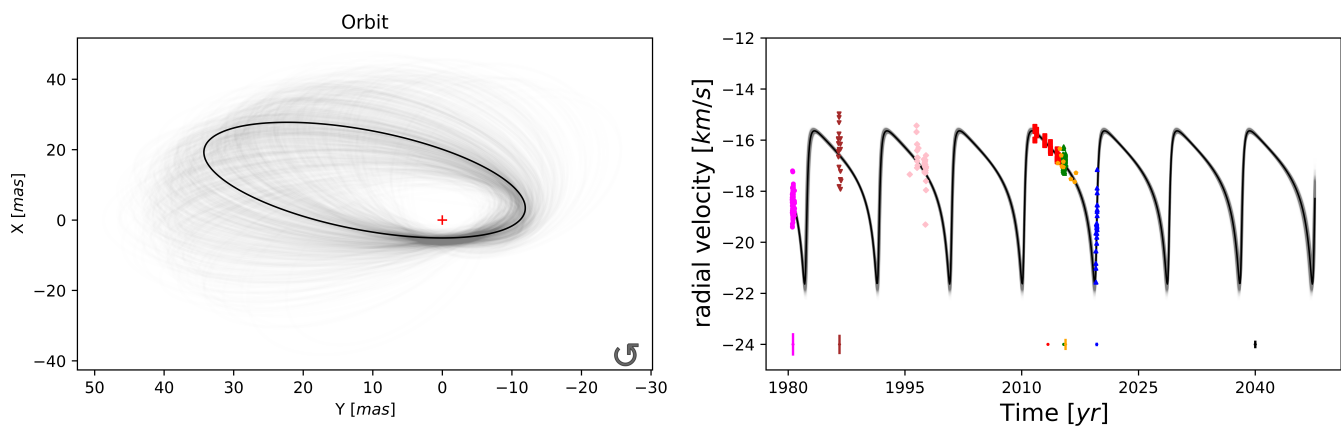}}
\end{center}
\caption{MAP point estimate projection of the a posteriori distribution for the estimated orbit (left panel) and the RV curve (right panel) for the SB1 system of $\delta$ Cep. The dark line corresponds to the MAP, and the light gray line shows the whole distribution. In the right panel, the radial velocity measurements are indicated with the same colors as in Figs. \ref{fig_wocor} and \ref{fig_wicor}. The mean uncertainties corresponding to each dataset are indicated at the bottom of the figure. The \citet{barnes05} dataset (shown with light pink diamonds) does not provide uncertainties. At the bottom right of the figure, we indicate in black the mean homogeneous uncertainty of 0.15 $\kms$ that we used for all dataset measurements to ensure a Bayesian fit.}
\label{fig_fit}
\end{figure*}

\begin{figure}[htbp]
\begin{center}
\resizebox{0.7\hsize}{!}{\includegraphics[clip=true]{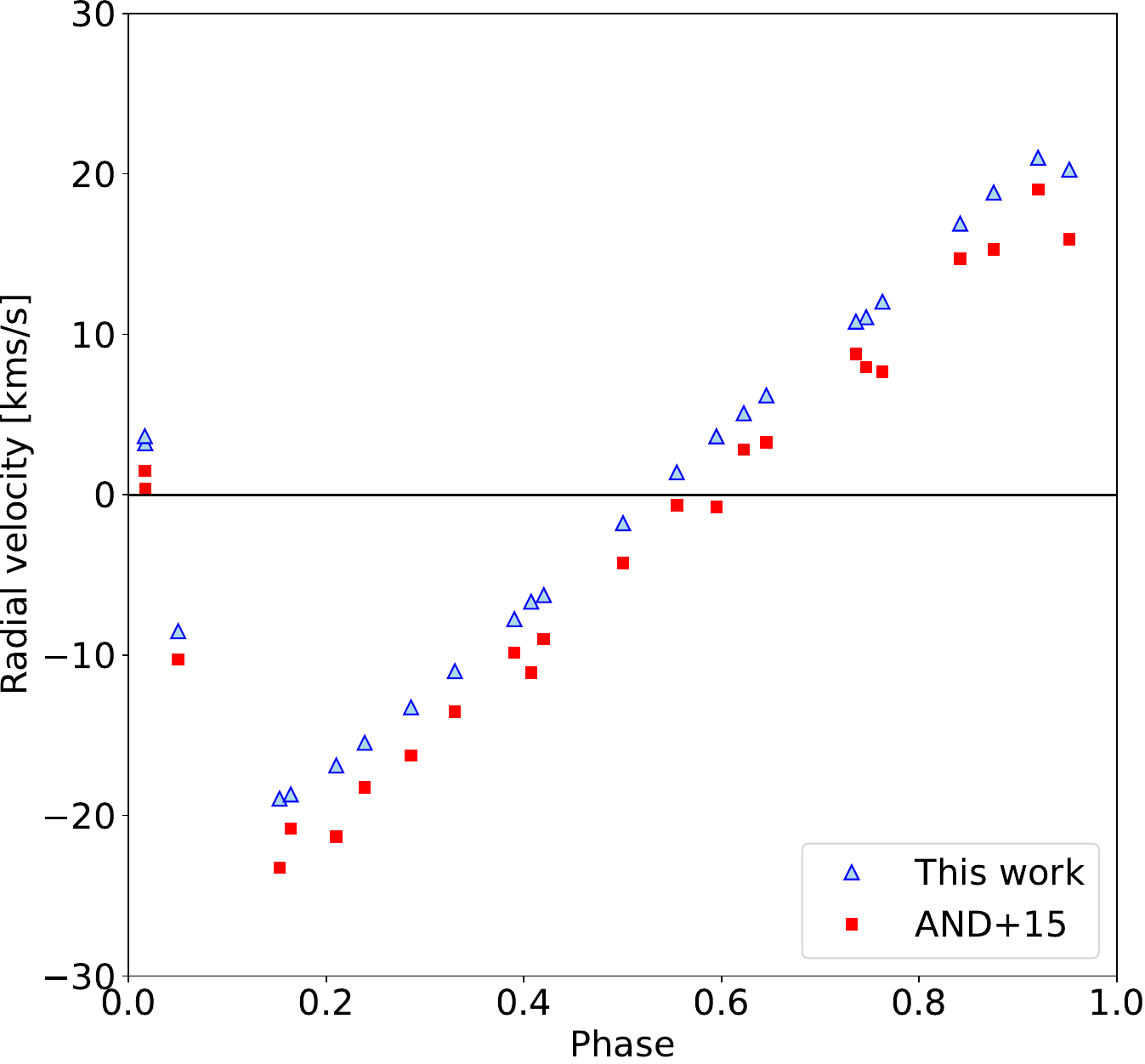}}
\end{center}
\caption{Final corrected RV$\mathrm{cc-g}$ curve of the HARPS-N data of $\delta$ Cep presented in this paper (blue triangles). For comparison, we also plot the corrected radial velocities determined when using the spectroscopic orbital parameters as found by \citet{anderson16} (red squares). }
\label{fig_final}
\end{figure}

The marginal posterior distributions of the orbital parameters are presented in Fig. \ref{fig_MAP}. The maximum a posteriori probabilities (MAPs) as derived from  the Bayesian inference in the multi-parameter space are indicated for each parameter with a vertical red line and are listed in Table \ref{Tab_res}. The horizontal black line shows the 95\% highest density probability intervals (HDPIs). The HDPI values are indicated in the figure and are also reported in Table \ref{Tab_res}. The left panel of Fig. \ref{fig_fit} shows the maximum probability visual estimated orbit (black line) together with the whole distribution of possible orbits (light gray lines). The right panel  shows the same, but for the maximum probability of the orbital radial velocity curve, with the spectroscopic measurements presented in Sect. \ref{s_data} overlaid. We also indicate at the bottom of the figure the mean uncertainties associated with each dataset. The scatter of old data (before the year 2000) compared to the orbital model appears larger than the individual uncertainties. This might be due to the fact that our template model of the pulsation of the star, based on the recent data from \citet{anderson15}, is not totally adapted to these old datasets because of uncorrected residuals in the secular period variation of the star \citep{Cso22}.  Using the maximum probability model (black line in the right panel of Fig. \ref{fig_fit}), we can correct  all  the spectroscopic radial velocity measurements in our sample for the orbital velocity of the Cepheid component (v$_\mathrm{orb}$). The derived corrected radial velocities are plotted in the left panel of Fig. \ref{fig_wicor}, and the residuals compared to our reference pulsation velocity template are shown as a function of the pulsation phase and BJD in the middle and right panels, respectively. The v$_\mathrm{orb}$ in the case of our new HARPS-N data from 2019 are indicated in Table \ref{Tab_log}, and the final corrected radial velocity curve is plotted in Fig. \ref{fig_final}. 

As shown by Table \ref{Tab_res} and in Fig. \ref{fig_MAP}, the orbital parameters are relatively well constrained. \citet{anderson16} found an orbital period of about $P_\mathrm{orb}=6.028 \pm 0.016$ years (median value of the distribution of probability), while we found a MAP of $P_\mathrm{orb}=9.32$ years. The two periods remain in the ratio 2:3, and the correct count of the elapsed cycles is not easy to determine when the rapid radial velocity variability due to an eccentric orbit is confined to a small phase interval and the observed radial velocity curve shows large gaps. In this respect, our 2019 data are particularly constraining since they cover the quick ascending branch (see the right panel of Fig. \ref{fig_fit}).  The derived radial velocity curve of $\delta$ Cep we obtain is of very good quality (Fig.~\ref{fig_final}). For comparison, in Fig.~\ref{fig_final} we also plot the corrected radial velocity when using the spectroscopic orbital parameters as found by \citet[see the red squares]{anderson16}. As an additional test, we arbitrarily considered offsets of $\pm 0.3$ \kms  on the different datasets used in this study, including our 2019 HARPS data, to simulate potential systematics or drifts in time between the different spectrographs. We find consistent MAP parameters (i.e., consistent within their uncertainties).  The rms of the residuals that we obtain using the orbital parameters we found is  about $0.4\,$km/s (Fig.~\ref{fig_wicor}, middle and right panels).


\section{Conclusion}\label{s_conclusion}

Using our latest HARPS-N spectroscopic dataset, from 2019, as well as data from the literature, we derived the orbital parameters of the SB1 binary system of the prototype of classical Cepheids, $\delta$~Cep. This allowed us to extract the radial velocity curve associated with the pulsation of $ \delta$~Cep, which is crucial to continuing our study of the dynamical structure of this star and, in particular, the Baade-Wesselink projection factor for Cepheids. According to our results, the system is now just after the quadrature in terms of orbital velocity, and securing new data in the near future will certainly help in confirming the orbital solution of the system. Furthermore, the fourth \textit{Gaia} data release will provide the epoch astrometric positions of the system, which will constrain the orientation of the orbit on the sky. As discussed in \citet{anderson16}, the discovery and characterization of the inner companion of $\delta$ Cep is important and should be investigated in the coming decade. 

\begin{acknowledgements}
The observations leading to these results have received funding  from the European Commission's Seventh Framework Programme (FP7/2013-2016)  under grant agreement number 312430 (OPTICON). The authors thank the GAPS observers F.~Borsa, L.~Di Fabrizio, R.~Fares, A.~Fiorenzano, P.~Giacobbe, J.~Maldonado, and G.~Scandariato.  This research has made use of the SIMBAD and VIZIER\footnote{Available at http://cdsweb.u- strasbg.fr/} databases at CDS, Strasbourg (France), and of the electronic bibliography maintained by the NASA/ADS system.  WG gratefully acknowledges financial support for this work from the BASAL Centro de Astrofisica y Tecnologias Afines (CATA) PFB-06/2007, and from the Millenium Institute of Astrophysics (MAS) of the Iniciativa Cientifica Milenio del Ministerio de Economia, Fomento y Turismo de Chile, project IC120009. 
WG also acknowledges support from the ANID BASAL project ACE210002.
Support from the Polish National Sci-
ence Center grant MAESTRO 2012/06/A/ST9/00269 and DIR-WSIB.92.2.2024 grants of the Polish Minstry of Science and Higher Education is also acknowledged. AG acknowledges the support of the Agencia Nacional de Investigaci\'on Cient\'ifica y Desarrollo (ANID) through the FONDECYT Regular grant 1241073. 
The authors acknowledge the support of the French Agence Nationale de la Recherche (ANR), under grant ANR-23-CE31-0009-01 (Unlock-pfactor) and the financial support from ``Programme National de Physique Stellaire'' (PNPS) of CNRS/INSU, France. A.G. acknowledges support from the ANID-ALMA fund No. ASTRO20-0059. B.P. gratefully acknowledges support from the Polish National Science Center grant SONATA BIS 2020/38/E/ST9/00486.  This work has made use of data from the European Space Agency (ESA) mission {\it Gaia}, processed by the {\it Gaia} Data Processing and Analysis Consortium (DPAC). Funding for the DPAC has been provided by national institutions, in particular the institutions participating in the {\it Gaia} Multilateral Agreement. The research leading to these results  has received funding from the European Research Council (ERC) under the European Union's Horizon 2020 research and innovation program (projects CepBin, grant agreement 695099, and UniverScale, grant agreement 951549). 
This work has made use of data from the European Space Agency (ESA) mission
{\it Gaia} (\url{https://www.cosmos.esa.int/gaia}), processed by the {\it Gaia}
Data Processing and Analysis Consortium (DPAC,
\url{https://www.cosmos.esa.int/web/gaia/dpac/consortium}). Funding for the DPAC
has been provided by national institutions, in particular the institutions
participating in the {\it Gaia} Multilateral Agreement.

\end{acknowledgements}

\bibliographystyle{aa}  
\bibliography{bibtex_nn} 

\break

\end{document}